\documentclass[nofootinbib,superscriptaddress,twocolumn]{revtex4}

\usepackage{amsmath}
\usepackage{amsfonts}
\usepackage{times}
\usepackage{graphicx}

\begin{document}

\title{Testing quantum-spacetime relativity with gamma-ray telescopes}

\author{Giovanni AMELINO-CAMELIA}
\affiliation{Dipartimento di Fisica, Universit\`a di Roma ``La Sapienza", P.le A. Moro 2, 00185 Roma, EU}
\affiliation{INFN, Sez.~Roma1, P.le A. Moro 2, 00185 Roma, EU}

\author{Antonino MARCIAN\`O}
\affiliation{Dipartimento di Fisica, Universit\`a di Roma ``La Sapienza",
 P.le A. Moro 2, 00185 Roma, EU}
 \affiliation{ ``Centre de Physique Th\'eorique'', Case 907 Luminy, 13288 Marseille, EU}

\author{Marco MATASSA}
\affiliation{Dipartimento di Fisica, Universit\`a di Roma ``La Sapienza",
P.le A. Moro 2, 00185 Roma, EU}

\author{Giacomo ROSATI}
\affiliation{Dipartimento di Fisica, Universit\`a di Roma ``La Sapienza", P.le A. Moro 2, 00185 Roma, EU}
\affiliation{INFN, Sez.~Roma1, P.le A. Moro 2, 00185 Roma, EU}

\begin{abstract}
\noindent Observations of gamma-ray bursts
are being used to test for a momentum dependence of the speed of photons,
partly motivated by preliminary results reported in analyses of some quantum-spacetime
scenarios.
The relationship between time of arrival, momentum of the photon and redshift of the source
which is used for these purposes assumes a ``breakdown" of relativistic symmetries,
meaning that it is
a preferred-frame scenario which
 does not
satisfy the  Relativity Principle.
The alternative hypothesis of a ``deformation" of relativistic symmetries,
which preserves the Relativity Principle by adopting
 deformed  laws of  relativistic transformation between
observers, could not so far be tested in gamma-ray-burst
observations because it was not known how to formulate it
in expanding  spacetimes.
We here provide such a formulation, and we find that also for the symmetry-deformation
scenario the analysis of gamma-ray-burst data take us very close to the desired ``Planck-scale sensitivity".
\end{abstract}

\maketitle

\baselineskip11pt plus .5pt minus .5pt

\noindent The famous second postulate of relativity,
the one that in high-school textbooks is simply described as a law of
constancy of the speed of light,
affects the observations performed by a gamma-ray telescope in a somewhat
subtle way.
A slight complication is due to the
expansion of our Universe:
within classical general relativity
the speed of light takes its
ideal value $c~(\simeq 2.998 \cdot 10^8m/s)$
only when the effects of ``spacetime curvature", such as the
Universe expansion, can be neglected.
For most Earth-bound experiments all such effects are indeed negligible, but
the gamma rays emitted by distant astrophysical objects
travel for times long enough
that the expansion of the Universe is tangible. And essentially
 the expansion causes inbound light to be partly dragged away from us, resulting in an effective
 speed which is smaller than $c$.

For classical spacetimes
 there is of course a robust and well-established description of the interplay
between speed of light and a possible expansion of spacetime,
and this description properly also factors in
the first relativity postulate, the ``Relativity Principle"
which requires the
equivalence of the laws of physics in all frames of reference.
But amusingly
in some areas of quantum-spacetime research
 for the last decade we only managed to characterize
some relativistic properties at the level of
a high-school-textbook description, without the ability
 to account for the Universe expansion.
This limitation in particular affects those models
of non-expanding quantum spacetimes (see, {\it e.g.}, Refs.~\cite{kpoinap,gacmajid,gampul,urrutiaPRL,aurelio,leekodama})
which have provided motivation for the study of the hypothesis
of momentum dependence
of the speed of photons.

The non-constancy of the speed of light found in some flat/non-expanding quantum spacetimes
definitely violates the second Einstein postulate, and it also
became quickly clear~\cite{gacdsrIJMPD2002,gacdsrPLB2001,kowadsr,leedsrPRL,dsrnature,jaconature} that
the nature and strength of the implications of this unexpected momentum dependence are affected
rather strongly by the fate of the first postulate, the postulation
of the ``Relativity Principle" of equivalence of reference frames.
Two alternative scenarios emerged. On one side there is
the LSB scenario\footnote{This is also described in the literature as
the LIV scenario (Lorentz Invariance Violation).} (Lorentz Symmetry Breakdown),
a ``quantum-gravity ether scenario" which
renounces~\cite{gampul,urrutiaPRL,grbgac} also to the validity
of the first postulate, and therefore ends up formulating fully nonrelativistic laws
with explicit reference to a ``preferred frame".
The alternative possibility is a ``deformation" of relativistic symmetries,
in the sense of the ``Doubly-Special Relativity"~\cite{gacdsrIJMPD2002,gacdsrPLB2001,kowadsr,leedsrPRL,dsrnature}
(DSR) scenario, which
 insists on a relativistic formulation
of laws, as required by the first postulate, and confines
all revisions of relativity to revisions of the second postulate.

The LSB/quantum-gravity-ether scenario is somewhat simpler from a technical perspective,
since one is spared the need of enforcing logical compatibility with the
relativity of frames of reference. But, even with this simplification,
it was not straightforward to
find a description
of the momentum dependence of the speed of photons that would be compatible
with the expansion of the Universe.
After a few steps~\cite{grbgac,ellisspeedold,piranspeed,ellisspeednew}
of gradual improvement,
 a consensus~\cite{piranspeed,ellisspeednew}
was reached on a formulation, which is centered on
a description of the speed of photons in a spacetime with scale factor $a$,
which in particular in conformal coordinates takes the form
\begin{eqnarray}
v_\gamma \simeq c \left( 1 - \lambda_{\textrm{\scriptsize{LSB}}}\frac{|p|}{a(\eta)} \right)\,,
 \label{vJOCCY}
\end{eqnarray}
where $p$ is the momentum\footnote{ We denote by $p$ the covariant spatial momentum ``$p_x$",
which is a conserved quantity in FRW spacetimes.
The contravariant spatial momentum ``$p^x$" is of course not conserved, but its scaling
along the worldline of a particle
is governed by the simple law $a^{-2}(\eta)p$.} of the photon and
 $\eta$ is the ``conformal time", related to the ``cosmological time" $t$
by $\eta = \int {dt}/{a(t)}$.

Most applications of this scenario were centered on
the associated prediction of
a difference in (cosmological-)times of arrival $\Delta t$ of two
photons, with momentum difference $\Delta p$,
emitted simultaneously by a source
at redshift $z$:
\begin{eqnarray}
\Delta t = \frac{1}{H_0} \lambda_{\textrm{\scriptsize{LSB}}} \Delta p\,  \int_0^z \frac{ (1+z') dz' }{\sqrt{ \Omega_m (1+ z')^3 + \Omega_\Lambda} }
~,
\label{dtzJOCCY}
\end{eqnarray}
where, as customary, we assumed expansion described by the current cosmological model,
with $H_0$ the Hubble constant ($\simeq 2.5 \cdot 10^{-18}s^{-1}$)
and the parameters $\Omega_\Lambda$ and $\Omega_m$
characteristic of the most successful ``$\Lambda$CDM cosmology"
($\Omega_\Lambda \simeq 0.73$, $\Omega_m \simeq 0.27$).

Over these two years past since a consensus~\cite{piranspeed,ellisspeednew}
was reached on this reformulation
for expanding spacetimes
of the original LSB proposal first given in Ref.~\cite{grbgac}
(for non-expanding spacetimes),
the prediction (\ref{dtzJOCCY}) was the unchallenged benchmark used in numerous
data analyses~\cite{magic2007,hessVGAMMA,fermiSCIENCE,FERMI090510,ellisLINEARinK} testing
the hypothesis of momentum dependence of the speed of photons.
One could therefore say that all these studies were exclusively testing
the hypothesis of ``quantum-spacetime {\underline{non}}-relativity".
The fact that there was no alternative DSR-relativistic picture
represents clearly the most significant limitation of the scope
of these studies.
We here set the stage for improving very significantly this state of affairs
by reporting
success in handling the severe technical challenges~\cite{kodadsr,triply,mignemiDSRSITTER,qds}
 that must be met
for a truly relativistic (no preferred frame)
description of momentum dependence of the speed of photons in an expanding spacetime.
Details will be provided in a longer companion paper~\cite{inprepNOI}.
Here we summarize the most intuitive aspects of our findings.
And for these purposes it is useful to take as starting point
the Eqs.~(\ref{vJOCCY}), (\ref{dtzJOCCY}) of the much-studied LSB proposal.

We wrote (\ref{vJOCCY}) in conformal coordinates because
this choice of coordinates of course
facilitates the comparison with results obtained for non-expanding spacetimes.
The DSR frameworks which we intend to adapt here
to spacetime expansion predict for non-expanding ($a=1$) spacetimes
\begin{eqnarray}
v_\gamma \Big|_{\textrm{\scriptsize{a=1}}}  \simeq c \left( 1 - \lambda_{\textrm{\scriptsize{DSR}}} |p| \right) ~.
 \label{vDSRminko}
\end{eqnarray}
Of course, also in the DSR case the
description of the momentum dependence of the speed
of photons requires the introduction of a length scale,
which we denote by $\lambda_{\textrm{\scriptsize{DSR}}}$.
But compatibility with the first relativity postulate imposes that
this length scale is an invariant, with the same value in all reference frames.
The main non-relativistic feature of
(\ref{vJOCCY})  is instead implicitly codified in the
 properties of
the length scale $\lambda_{\textrm{\scriptsize{LSB}}}$,
which is introduced in the LSB scenario
as a scale
that acquires different values~\cite{grbgac,gampul,urrutiaPRL,piranspeed,gacflavtsviuri}
 in different frames of reference.

 The observer-independence of  $\lambda_{\textrm{\scriptsize{DSR}}}$
 was the main objective of our analysis,
 but we were also somewhat concerned about another aspect
 of (\ref{vJOCCY}), which is the dependence of $v_\gamma$ on $a(\eta)$:
it implies that in such ``quantum-gravity ether" pictures of quantum spacetime the adoption
of conformal coordinates
 does not actually achieve the usual goal of factoring out the
Universe expansion for photons.
We should stress that there is no {\it a priori} reason to motivate this feature.
Our present (admittedly limited) understanding of the quantum-spacetime concept
does not suggest in any way that the speed of photons should, unlike what happens
in the classical spacetime case ($\lambda_{\textrm{\scriptsize{LSB}}} \rightarrow 0$), retain
a (conformal-)time dependence even in conformal coordinates.

Our search of a DSR framework that
would be compatible with spacetime expansion was
eventually successful, but only after exploiting the simplification
of our task that was provided by the adoption of conformal coordinates.
Our intuition was that in conformal coordinates we should be able to
use rather directly the lessons learned by applications of the DSR concept
in non-expanding spacetimes.
And it is noteworthy that some of the most used formulations of such DSR theories
are Hamiltonian theories which add to the familiar Hamiltonian describing
a free particle in Minkowski spacetime ($a=1$) the
correction term $\lambda_{\textrm{\scriptsize{DSR}}} p^2/2$:
\begin{eqnarray}
\mathcal{H}_{\textrm{\scriptsize{DSR}}} \Big|_{\textrm{\scriptsize{a=1}}}
=\sqrt{p^2+m^2} -\frac{\lambda_{\textrm{\scriptsize{DSR}}}}{2} p^2
~.
\label{hamilPOINC}
\end{eqnarray}
The rather strong connection between the natural coordinates of Minkowski spacetime
and the conformal coordinates of an expanding (FRW) spacetime
led us to the proposal of adding the
correction term $\lambda_{\textrm{\scriptsize{DSR}}} p^2/2$
also to the Hamiltonian of a free particle in a classical FRW spacetime:
\begin{eqnarray}
\mathcal{H}_{\textrm{\scriptsize{DSR}}}=\sqrt{p^2+m^2 a^2(\eta)} -\frac{\lambda_{\textrm{\scriptsize{DSR}}}}{2}p^2
~.
\label{hamilDESITTER}
\end{eqnarray}
One easily finds~\cite{inprepNOI} that (\ref{hamilDESITTER}) produces worldlines
governed by the following description of the speed of photons
in conformal coordinates for an expanding spacetime:
\begin{eqnarray}
v_\gamma \simeq c (1 - \lambda_{\textrm{\scriptsize{DSR}}} \, |p| )
~,
 \label{vNOI}
\end{eqnarray}
 which is (conformal-)time independent.

 And most importantly we find~\cite{inprepNOI} that this theory
 governed by the Hamiltonian $\mathcal{H}_{\textrm{\scriptsize{DSR}}}$
 of (\ref{hamilDESITTER}) has the length scale
$ \lambda_{\textrm{\scriptsize{DSR}}}$  with the status
of a relativistic invariant.
We find that ``locally" ({\it i.e.} in analyses local enough that the time dependence of
the scale factor can be neglected)
the equations of motion are covariant, and
the scale $\lambda_{\textrm{\scriptsize{DSR}}}$ is invariant, under a rather
typical~\cite{gacdsrIJMPD2002,leedsrPRL,gacdsr2010review}
DSR realization of relativistic symmetries, centered on the following deformed representation of boosts
\begin{eqnarray}
  {\cal N} = x \partial_\eta + \eta \partial_x + \frac{\lambda_{\textrm{\scriptsize{DSR}}}}{2} x \partial_x^2- \lambda_{\textrm{\scriptsize{DSR}}} \, \eta \, \partial_\eta \partial_x
 ~.
 \label{boostPOINC}
\end{eqnarray}
And
for the de Sitter case, $a(\eta)=(1-H\eta)^{-1}$,
 we rather remarkably also find that globally the equations of motion are covariant, and
the scale $\lambda_{\textrm{\scriptsize{DSR}}}$ is invariant, under a
novel DSR realization of relativistic symmetries compatible with de Sitter
expansion, which is centered on the following deformed representation of boosts~\cite{inprepNOI}
\begin{eqnarray}
&{\cal N}&=x (1-H \eta) \partial_\eta+\left(\frac{1-(1-H\eta)^{2}}{2H}-\frac{H}{2}x^{2}\right)\partial_x + \nonumber\\
&+&\!\!\!\!\lambda_{\textrm{\scriptsize{DSR}}} \left( \frac{1+ H \eta }{2}\, x \,\partial_x - (1- H \eta)\, \eta\, \partial_\eta + \frac{H \eta}{2} \right) \partial_x\,.
 \label{boostDS}
\end{eqnarray}

We are therefore providing for the first time a fully relativistic theory with
momentum dependence of the speed of photons which is compatible with spacetime expansion,
and this finally allows one to test the deformed-symmetry scenario
with data gathered by gamma-ray telescopes.
The strategy is analogous to the one already in use for the LSB/ether-based
predecessor: one relies~\cite{grbgac}
on observations of sources that produce
short-duration but intense bursts of gamma-ray photons,
looking for indirect evidence of momentum-correlated
differences in the times of arrival of the photons.
If such momentum-dependent delays are found and
cannot be ascribed to differences in times of emission, then they would be
interpreted as a manifestation of a momentum-dependent speed of propagation.

While the strategy of the data analysis is analogous, there are two profound differences
between the LSB framework centered on (\ref{vJOCCY}) and our DSR framework centered
on (\ref{vNOI}).
The conceptually most significant difference between
(\ref{vJOCCY}) and (\ref{vNOI}) concerns the different status of the
scales $\lambda_{\textrm{\scriptsize{LSB}}}$
and $\lambda_{\textrm{\scriptsize{DSR}}}$, which we already stressed: the fact that our $\lambda_{\textrm{\scriptsize{DSR}}}$ is
absolutely invariant allows one to combine straightforwardly the results of all such data
analyses, whereas for $\lambda_{\textrm{\scriptsize{LSB}}}$ data obtained by different telescopes,
or even by the same telescope at different times (when, according to the ether picture,
the telescope has different velocity with respect to the ``preferred frame of reference"),
can be combined only in rather nontrivial way, only by carefully mapping all results
to the same (``preferred") frame of reference.
This is the core difference between relativistic and nonrelativistic theories,
but quantitatively, at least for the present generation of telescopes,
it carries very little weight: taking, for example, the natural frame for the description
of the cosmic microwave background as the preferred frame, one finds that
the speeds of our telescopes in that frame
are all very small with respect to $c$,
and this implies \cite{gacflavtsviuri} that also the
manifestations of the non-invariance of $\lambda_{\textrm{\scriptsize{LSB}}}$ are negligibly small at present.
Since speeds of order $\sim 10^{-4}c$ are not uncommon for space telescopes
one can envision this issue to come into play  (if and) when the length scale
characteristic of the momentum dependence of the speed of light is measured
with precision of 1 part in $10^4$. This is a possibility which is presently  beyond our
reach, but intriguingly it is not inconceivable for a not-so-distant future.

The second difference
between (\ref{vJOCCY}) and our (\ref{vNOI})
is less significant conceptually, since it is not an explicit manifestation
of the relativistic nature of our proposal,
but does affect the analysis in a way that is quantitatively significant.
This is the fact already stressed above
that our analysis produces a law for the speed of photons
which adopting
conformal coordinates is time independent, whereas somewhat surprisingly the LSB/ether-based
result (\ref{vJOCCY}) does depend on conformal time, through the expansion factor $a(\eta)$.
As a result our DSR proposal is applicable also to the phenomenology of the early  Universe, where
(as stressed above) the popular LSB framework runs into concerning pathologies.
And more importantly, even for the presently most actively pursued opportunity, the one concerning
observations of bursts of gamma rays,
the difference between (\ref{vJOCCY}) and our (\ref{vNOI})
is very tangible for data analysis,
as shown in Figs.~1 and 2.

\begin{figure}[h!]
\includegraphics[width=0.48\textwidth]{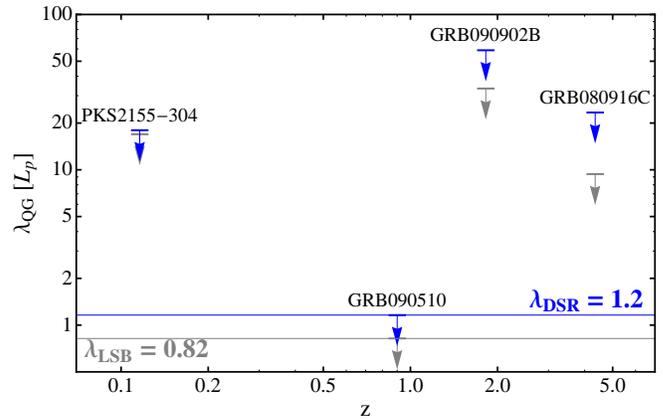}
\caption{Here we give an indication of the significance of the differences between the previously-used ether-based Eq.~(\ref{vJOCCY}) and our relativistic Eq.~(\ref{vNOI}),
by comparing the upper bounds on $\lambda_{\textrm{\scriptsize{LSB}}} $ (grey)
and $\lambda_{\textrm{\scriptsize{DSR}}}$ (blue)
that can be inferred from the four most significant relevant analyses of bursts of gamma rays,
reported in Refs.~\cite{FERMI090510,hessVGAMMA,fermiSCIENCE,ellisLINEARinK}.
Different sources
 are identified by their redshift, with the primary objective
of showing that data from sources at redshift up to $z \lesssim 1$ have quantitatively
similar implications for $\lambda_{\textrm{\scriptsize{DSR}}}$ and for $\lambda_{\textrm{\scriptsize{LSB}}}$, but
for $z$ significantly greater than $1$ (where however presently the quality of data
is not very good) the differences are rather sizeable.}
\end{figure}

In Fig.~1 we compare the implications of the same observations
 on our relativistic proposal (parameter $\lambda_{\textrm{\scriptsize{DSR}}}$)
 and on the previously-considered ether-based scenario (parameter $\lambda_{\textrm{\scriptsize{LSB}}}$).
The limits on $\lambda_{\textrm{\scriptsize{DSR}}}$ that can be derived from different
observations of bursts of gamma rays are described in units of
the Planck length $L_P$ ($\equiv \sqrt{ {\hbar G}/{c^3}} \simeq 1.6 \cdot 10^{-35} ~ m$),
which according to a few independent (but all semi-heuristic, see, {\it e.g.}, Refs.~\cite{garay,adler}) arguments should be within one or two orders of magnitude of
the characteristic scale of quantum-gravity/quantum-spacetime effects
(so we should expect $\lambda_{\textrm{\scriptsize{DSR}}}$ to be within
one or two orders of magnitude of  $L_P$).
The main message that
our Fig.~1 intends to convey is that our analysis provides the basis
for testing a relativistic (no-ether)
description of momentum dependence of the
speed of photons with sensitivity which is already presently at the ``Planck-length
sensitivity" level.

We find that the most stringent limit on $\lambda_{\textrm{\scriptsize{DSR}}}$
is inferred from the observation of gamma-ray burst GRB090510 by the Fermi space telescope.
It was already shown in Ref.~\cite{FERMI090510} that for GRB090510,
from a source at redshift $z \simeq 0.9$, the data analysis could establish that
any time delay due to momentum-dependent speeds of photons was not greater
than $30\, ms$, at momenta of $\sim 1 GeV$.
This upper bound on momentum-dependence-induced time-of-arrival
differences was shown~\cite{FERMI090510} to imply a limit of $\lambda_{\textrm{\scriptsize{LSB}}} < 0.82 L_P$
for the ether-based case. We find that for our relativistic scenario the
same data imply $\lambda_{\textrm{\scriptsize{DSR}}} < 1.2 \, L_P = 1.9 \cdot 10^{-35}m$,
and is the best present limit
on $\lambda_{\textrm{\scriptsize{DSR}}}$.

The fact that the data on GRB090510 imply a limit on $\lambda_{\textrm{\scriptsize{LSB}}}$
and a limit on $\lambda_{\textrm{\scriptsize{DSR}}}$ with not much difference between them
 shows that even at
redshifts as large as $z \simeq 1$ the differences between
our new DSR/relativistic result (\ref{vNOI})
and its LSB/ether-based counterpart (\ref{vJOCCY}),
while tangible, are not very large.
Fig.~1 also considers the case of the observation by the HESS telescope~\cite{hessVGAMMA}
of a flare from the Active
Galactic Nucleus PKS2155-304, whose implications
for $\lambda_{\textrm{\scriptsize{LSB}}}$
and for $\lambda_{\textrm{\scriptsize{DSR}}}$ are quantitatively the same (within $5 \%$),
because PKS2155-304 is at the relatively small redshift of $z\simeq 0.116$.
But the cases in Fig.~1 which are at redshift significantly greater than~1 (GRB090902B and GRB080916C,
observed by the Fermi telescope\footnote{The ``time-delay information content"
of the observation of GRB080916C has been investigated
in detail by the Fermi collaboration~\cite{fermiSCIENCE},
while the corresponding
analysis of GRB090902B has been so far only preliminarily reported in Ref.~\cite{ellisLINEARinK}.})
illustrate the fact that at large redshifts the effects
of our relativistic model are significantly different from
the ones of the ether-based model.

Fig.~2 contributes from a different angle to the same assessment of the dependence on redshift of
the differences between the much studied LSB scenario with $\lambda_{\textrm{\scriptsize{LSB}}}$
and our novel DSR scenario. With Fig.~2 we point out that, if data of quality roughly comparable
with the observation of GRB090510 is obtained from sources at somewhat higher redshift
(say, redshifts in the range from $z\simeq 2$ to $z\simeq 5$) the difference between
the LSB scenario and our DSR scenario will be observably large.

\begin{figure}[h!]
\includegraphics[width=0.49\textwidth]{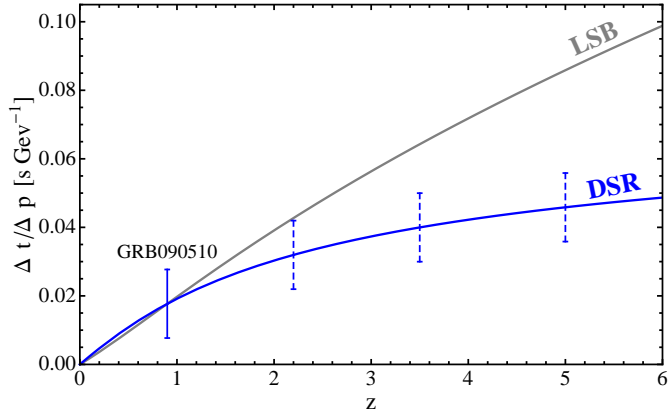}
\caption{Comparison of the dependence of $\Delta t/\Delta p$ on redshift for
our DSR (blue line) and the LSB (grey line) scenario, in the case when the
parameters $\lambda_{\textrm{\scriptsize{DSR}}}$  and   $\lambda_{\textrm{\scriptsize{LSB}}}$
 are adjusted in such a way that the two predictions for $\Delta t/\Delta p$
 are the same at $z \simeq 0.9$ (the redshift of GRB090510).
 At large redshift the differences are significant enough that one could plausibly
 discriminate experimentally between the two scenarios. This however would require
 much better control of systematic errors than we have presently (see later comments to Fig.~3)
 and would also require that the statistical errors for data analyses
 on high-redshift sources (hypothetically drawn
 as dashed blue vertical intervals) are kept comparable to the statistical error
 found in the analysis of GRB090510 (solid blue vertical interval).}
\end{figure}

Fig.~3 offers another perspective on presently-available data
centered on the fact that,
upon re-expressing our conformal-coordinate formulation in terms of
cosmological time and redshift
measurements, one finds that our
result (\ref{vNOI}) leads to the prediction
of a linear relation
$$\Delta t = \frac{1}{H_0} \lambda_{\textrm{\scriptsize{DSR}}} \Delta p\, K_{\textrm{\scriptsize{DSR}}}(z)$$
between the difference in times of arrival $\Delta t$ of two simultaneously-emitted
photons, with momentum difference $\Delta p$, and the following function of the
redshift $z$:
\begin{eqnarray}
K_{\textrm{\scriptsize{DSR}}}(z)=  \int_0^z \frac{  dz' }{\sqrt{ \Omega_m (1+ z')^3 + \Omega_\Lambda} }
~
\end{eqnarray}
(where we assumed again $H_0 \simeq 2.5 \cdot 10^{-18}s^{-1}$,
$\Omega_\Lambda \simeq 0.73$, and $\Omega_m \simeq 0.27$).

\begin{figure}[h!]
\includegraphics[width=0.48\textwidth]{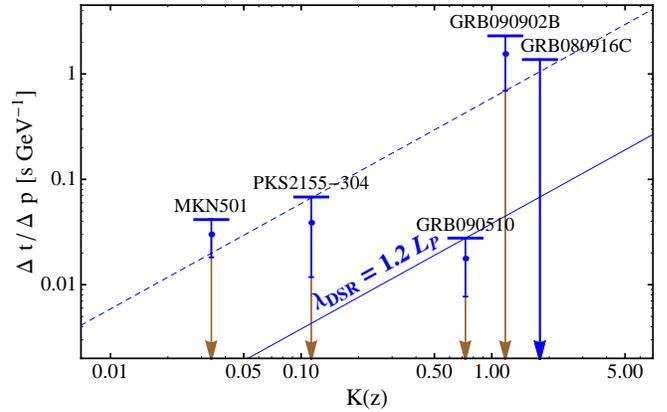}
\caption{Here we mainly intend to emphasize that in order to determine from data
the value of our
relativistic scale $\lambda_{\textrm{\scriptsize{DSR}}}$ one can exploit
the linear dependence on $K_{\textrm{\scriptsize{DSR}}}(z)$
of the $\Delta t/\Delta p$ (expected delay of arrival versus momentum difference
of simultaneously emitted photons). We show two such linear relations,
one (solid line) for the value of $\lambda_{\textrm{\scriptsize{DSR}}}=1.2 L_P$,
which we establish as the best present limit on $\lambda_{\textrm{\scriptsize{DSR}}}$ (from data~\cite{FERMI090510}
on GRB090510,
at redshift $z \simeq 0.9$, {\it i.e.}~$K_{\textrm{\scriptsize{DSR}}}(z) \! \simeq \! 0.7$),
and one for $\lambda_{\textrm{\scriptsize{DSR}}}=18 L_P$,
which is the second best limit (from data~\cite{hessVGAMMA} on PKS2155-304).
Following the thesis put forward in the recent Ref.~\cite{ellisLINEARinK}, the figure suggests that
in some cases the mere analysis of statistical errors (blue vertical intervals) could naively invite
a positive/discovery interpretation, but (as suggested by the brown vertical lines) essentially
we still have no control on a peculiar systematic error that affects these analyses,
which is due to our poor knowledge of the mechanisms that produce the structure of
the bursts at the source (the preliminary indications of correlation between
momentum of the photon and time of its detection at the telescope are likely to be entirely due
to properties of the source).}
\end{figure}

Another key difference between our new DSR/relativistic framework and
the previous LSB/ether-based picture concerns the stability of photons.
Our formalization of deformed relativistic symmetries is powerful enough
that one can then easily prove that
the photon is stable (for example the requirements of DSR-relativistic kinematics in
expanding spacetime are such that the process $\gamma \rightarrow e^+ e^-$ is never allowed~\cite{inprepNOI}).
The description of DSR-relativistic kinematics in
expanding spacetime was one of the key missing pieces of previous
attempts of DSR formulation~\cite{kodadsr,triply,mignemiDSRSITTER,qds}.
And it marks a very significant difference with respect to LSB/quantum-gravity-ether
theories where photon decay can be a significant phenomenological concern~\cite{jlm}.

The two phenomenological pictures we have been comparing, the previously developed LSB
scenario and our novel DSR scenario,
clearly present significant differences. And we feel that both of them deserve intrinsic
interest, with or without quantum gravity, just because the importance
of the development of novel
Lorentz-symmetry test theories must be assessed as proportional to the pivotal role
that exact/classical Lorentz symmetry enjoys in our present formulation of the
fundamental laws of Nature.
Still it is of course interesting to ponder on the implications
that these two scenarios could have specifically for quantum-gravity research.
Because of the complexity of the relevant quantum-spacetime theories, at present
we only have relatively robust ``theoretical evidence"  (in some models~\cite{kpoinap,gacmajid,gampul,urrutiaPRL,aurelio,leekodama})
of modifications
of the energy-momentum dispersion relation for non-expanding spacetimes,
without any robust indication for or against the emergence of a preferred frame,
and without the ability to predict the implications of curvature/expansion of spacetime
for the relevant effects.

Our analysis provides guidance for future works on the theory side attempting
to establish more robustly the links from models of quantum spacetime
and phenomenological pictures concerning the fate of Lorentz symmetry
at the Planck scale.
On the issue of the possible emergence (or DSR/non-emergence)
of a preferred frame there was already
a reference conceptual framework, discussed {\it e.g.} in Refs.~\cite{gacdsrIJMPD2002,leekodama}.
We here provided several new elements that could play an important
role in future analyses of these issues.
In particular we observed that
attempts to give sharper quantum-gravity motivation for
the previously developed LSB scenario
should look for evidence of a mechanism such that
 the adoption of conformal coordinates
 would not actually achieve the usual goal of factoring out the
Universe expansion for photons. The possibility of such a ``failure" of conformal coordinates
had never been raised before in the quantum-gravity literature and we feel
it can be of valuable guidance.

More insight can be gained by looking at the
specific conformal-time dependence
predicted by (\ref{vJOCCY}), for the previously developed LSB scenario.
If one takes at face value (\ref{vJOCCY}) then
the presence of the scale factor in the denominator
of the correction term produces, as shown in Fig.~4,
some rather puzzling results:
for some correspondingly small values of the scale factor
the world-lines described by (\ref{vJOCCY}) are rather pathological
and the energy of the photons takes imaginary values~\cite{inprepNOI}.
In order to improve on this state of affairs it appears natural to
invoke the presence of further correction terms\footnote{We gratefully acknowledge
valuable conversations with T. Piran concerning these additional correction terms.}
(to be added to (\ref{vJOCCY}))
of type, say,  $\lambda^2 p^2 a^{-2}$ and with sign and magnitude adjusted in such
a way to eliminate the unwanted features of the worldlines.
In turn this observation can provide  a key target for attempts to reproduce
the previously developed LSB scenario within some chosen quantum-gravity picture:
it appears that these attempts should first of all find evidence of a mechanism such
that the strength of quantum-gravity effects is governed by $p a^{-1}$,
which for a massless particle is the ``comoving energy" (energy in ``comoving coordinates", which are like
conformal coordinates but with
cosmological time rather than conformal time).

And it is noteworthy that such a mechanism would be easily distinguished
from the type of mechanism that a quantum-gravity theory should host in order
to provide support for our novel DSR picture,
where the magnitude of quantum-gravity effects is governed by the conserved charge $p$.
In our DSR picture the significance of quantum-gravity effects would be weighed by comparing
the $p$ that labels the worldline and the (observer-independent) scale $1/\lambda_{\textrm{\scriptsize{DSR}}}$:
some worldlines could be described neglecting the quantum-gravity effects completely,
while other worldlines (with $p$ comparable to $1/\lambda_{\textrm{\scriptsize{DSR}}}$)
would be severely affected by the quantum-gravity effects.
In the previously developed LSB scenario, since the significance of quantum-gravity effects
would be determined by comparing $p a^{-1}$ to the (observer-dependent)
scale $1/\lambda_{\textrm{\scriptsize{LSB}}}$,
one would instead typically find that different portions of the same worldline
are affected differently by the quantum-gravity effects: portions where the scale factor
is not small would be largely unaffected, while portions where the scale factor is
very small would reflect quantum-gravity properties very strongly.

\begin{figure}[h!]
\includegraphics[width=0.44\textwidth]{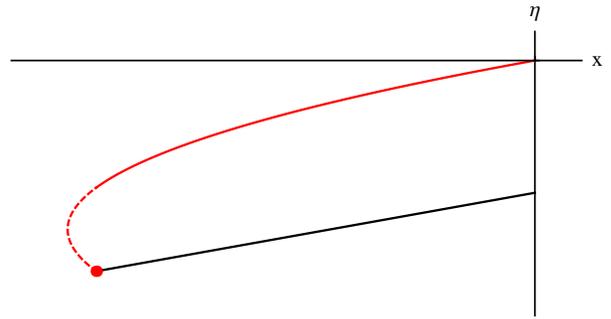}
\caption{We here illustrate qualitatively a previously unnoticed
feature of
the much-studied formalization of the LSB scenario for expanding spacetimes,
here described in Eqs.~(\ref{vJOCCY}), (\ref{dtzJOCCY}). For values of the conformal time
such that $a \ll 1$ one finds that the path of photons
of momentum $\sim a/\lambda$ (much below the Planck scale)
from a static source to a static telescope (both static in comoving coordinates
of course) starts off going in the ``wrong direction", and
gets going
in the ``right direction"
only after some time, when the scale factor
 reaches larger values.  This is shown by the red line, which is also compared to the case (black line)
 of a photon emitted simultaneously by the same source with momentum $\ll a/\lambda$.
 During the time of the ``world-line anomaly" one also finds~\cite{inprepNOI}
that the energy is imaginary.}
\end{figure}

\newpage
The possibility that the differences we here exposed
could be exploited to discriminate  between
the much-studied LSB framework of Eqs.~(\ref{vJOCCY}), (\ref{dtzJOCCY})
and our novel DSR framework is certainly exciting.
Of course, the most likely scenario  is
the one with Einstein's relativity still prevailing, at least for now.
We assumed here effects suppressed only by one power of the Planck length,
because this is what is afforded by the sensitivity of presently-operating
gamma-ray observatories.
But from a quantum-gravity/quantum-spacetime perspective
the case of quadratic ($\lambda^2_{\textrm{\scriptsize{DSR}}}$)
suppression appears to be equally natural~\cite{ellisLINEARinK,piranNEUTRINO,gacSMOLIN2009},
and could
become accessible in a not-too-distant future,
exploiting~\cite{piranNEUTRINO} observations
of the ultra-high-energy neutrinos
that accompany~\cite{waxneutri} gamma-ray bursts.
The techniques we here discussed
for discriminating between DSR/relativistic frameworks
and LSB/ether-based frameworks should prove valuable
for that future challenge,
and are already a necessary resource
for fully exploiting the present
opportunities for challenging Einstein relativity
with corrections that depend linearly on the
 Planck length.

\section*{Acknowledgments}
\noindent
G.~A.-C. is supported in part by grant RFP2-08-02 from The Foundational Questions Institute (fqxi.org).\\
We gratefully acknowledge valuable conversations with F.~Fiore, G.~Gubitosi, F.~Mercati and L.~Smolin.

\end{document}